\documentclass[5p,sort&compress]{elsarticle}

\usepackage{lineno}
\usepackage{url}
\usepackage{bm}
\usepackage{amsmath}
\usepackage{graphicx}
\usepackage{SIunits}
\usepackage{xcolor}
\usepackage[colorlinks=true
,urlcolor=blue
,anchorcolor=blue
,citecolor=blue
,filecolor=blue
,linkcolor=blue
,menucolor=blue
,pagecolor=blue
,linktocpage=true
,pdfproducer=medialab
,pdfa=true
]{hyperref}

\modulolinenumbers[5]

\newcommand{\hsigma}{{\hat{\sigma}}}
\newcommand{\dd}{{\textrm{d}}}

\pdfoutput=1

\journal{Physics Letters B}

\begin{document}

\begin{frontmatter}

\title{Gluon fusion and $b\bar{b}$ corrections to $H W^+ W^- / H Z Z$
  production in the POWHEG-BOX}

\author{Julien Baglio}
\address{Institut f\"{u}r Theoretische Physik, Eberhard Karls
  Universit\"at T\"ubingen, Auf der Morgenstelle 14, D-72076
  T\"ubingen, Germany}
\ead{julien.baglio@uni-tuebingen.de}

\begin{abstract}
The study of the Higgs boson properties is one of the most important
tasks to be accomplished in the next years, at the Large Hadron
Collider (LHC) and at future colliders such as the Future Circular
Collider in hadron-hadron mode (FCC-hh), the potential 100 TeV
follow-up of the LHC machine. In this view the precise study
of the Higgs couplings to weak gauge bosons is crucial and requires as
much information as possible. After the recent calculation of the
next-to-leading order QCD corrections to the production cross
sections and differential distributions of a Standard Model Higgs
boson in association with a pair of weak bosons, matched with parton
shower in the {\tt POWHEG-BOX} framework, we present the gluon fusion
correction $g g\to H W^+_{} W^-_{} ( H Z Z)$ to the process $p p \to H
W^+_{} W^-_{} (H Z Z)$. This correction can be sizeable and amounts to
$+3\,\%$ ($+10\,\%$) in the $H W^+_{} W^-_{}$ process and $+5\,\%$
($+18\,\%$) in the $H Z Z$ process at the LHC (FCC-hh). We also
present the first study of the impact of the bottom--quark initiated
channels $b\bar{b}\to H W^+_{} W^-_{} / H Z Z$ and find that they
induce a significant $+18\,\%$ correction in the $H W^+_{} W^-_{}$
channel at the FCC-hh. We present results on total cross sections and
distributions at the LHC and at the FCC-hh.
\end{abstract}


\end{frontmatter}


\section{Introduction}
\label{sec:intro}

After the discovery of a Higgs boson with a mass of
$\sim$~\unit{125}{GeV} in the Run I of the Large Hadron Collider
(LHC) at CERN~\cite{Aad:2012tfa,Chatrchyan:2012ufa}, the study of its
properties has begun, in particular to test whether they deviate from
the predictions of the Standard Model (SM)
mechanism~\cite{Higgs:1964ia,Higgs:1964pj,Guralnik:1964eu}. The latest
results at 13 TeV still display a compatibility with the SM
hypothesis~\cite{ATLAS-CONF-2016-081,CMS-PAS-HIG-16-020,CMS-PAS-HIG-16-033}. Developing
the most exhaustive survey of possible deviations from the SM is thus
an important task. In this view the coupling between a Higgs boson and
weak bosons is a crucial part of this survey. The production of a
Higgs boson in association with a pair of weak gauge
bosons~\cite{Baillargeon:1993iw,Cheung:1993bm,Djouadi:2005gi,Baglio:2015wcg}
can be used to probe the Higgs gauge
couplings~\cite{Gabrielli:2013era}, which is also directly related to
the triple gauge bosons vertex~\cite{Corbett:2013pja}.

The next-to-leading order (NLO) QCD corrections to various $H+VV'$
processes at the LHC have now been calculated: $HW^+_{} W^-_{}$
production~\cite{Mao:2009jp,Alwall:2014hca,Baglio:2015eon}, $HW^\pm _{} Z$
production~\cite{Liu:2013vfu,Alwall:2014hca,Baglio:2015eon},
associated production with a massive gauge boson $W/Z$ and a
photon~\cite{Mao:2013dxa,Alwall:2014hca,Shou-Jian:2015sta}, and
finally $HZZ$ production~\cite{Alwall:2014hca,Baglio:2015eon}. The
matching with parton shower for all processes was done in the {\tt
  MadGraph5\char`_aMC@NLO} framework in 2014~\cite{Alwall:2014hca}
(including all loop-induced gluon fusion
contributions~\cite{Hirschi:2015iia}) and was also completed in 2015
in the {\tt POWHEG-BOX} framework~\cite{Frixione:2007vw,Alioli:2010xd}
for all processes except those involving a
photon~\cite{Baglio:2015eon}. The gluon fusion correction to $H W^+_{}
W^-_{}$, $g g\to H W^+_{} W^-_{}$, exists in the
literature~\cite{Mao:2009jp,Hirschi:2015iia} and amounts to $\sim
+4\,\%$ to the total cross section at the LHC for a fixed central
scale. The gluon fusion correction to $H Z Z$ was calculated in
Ref.~\cite{Hirschi:2015iia} and amounts to $+0.1$~fb to the total
cross section at the 13 TeV LHC for a dynamical central scale.

This Letter is a follow-up to Ref.~\cite{Baglio:2015eon} and completes
the picture in the {\tt POWHEG-BOX} framework by presenting the gluon
fusion corrections to $H W^+_{} W^-_{}$ and $H ZZ$ production in the
SM and their matching with parton shower. We will also present, for
the first time, a study of the impact of bottom--quark contributions
in the five-flavour scheme, $p p\to b \bar{b}\to H W^+_{} W^-_{} / H Z
Z$. We find that they induce significant corrections in particular in
the $H W^+_{} W^-_{}$ channel. The paper is organised as follows: in
Section~\ref{sec:tools} the calculation and the tools used are
presented, then in Section~\ref{sec:numres} the numerical results are
presented, both for the LHC and the Future Circular Collider in
hadron-hadron mode (FCC-hh), the potential machine which would follow
the LHC with an energy of 100 TeV. A short conclusion will be given in
Section~\ref{sec:conclusions}.

\section{Description of the calculation}
\label{sec:tools}

\begin{figure*}[t]
  \centering
  \includegraphics[scale=0.5]{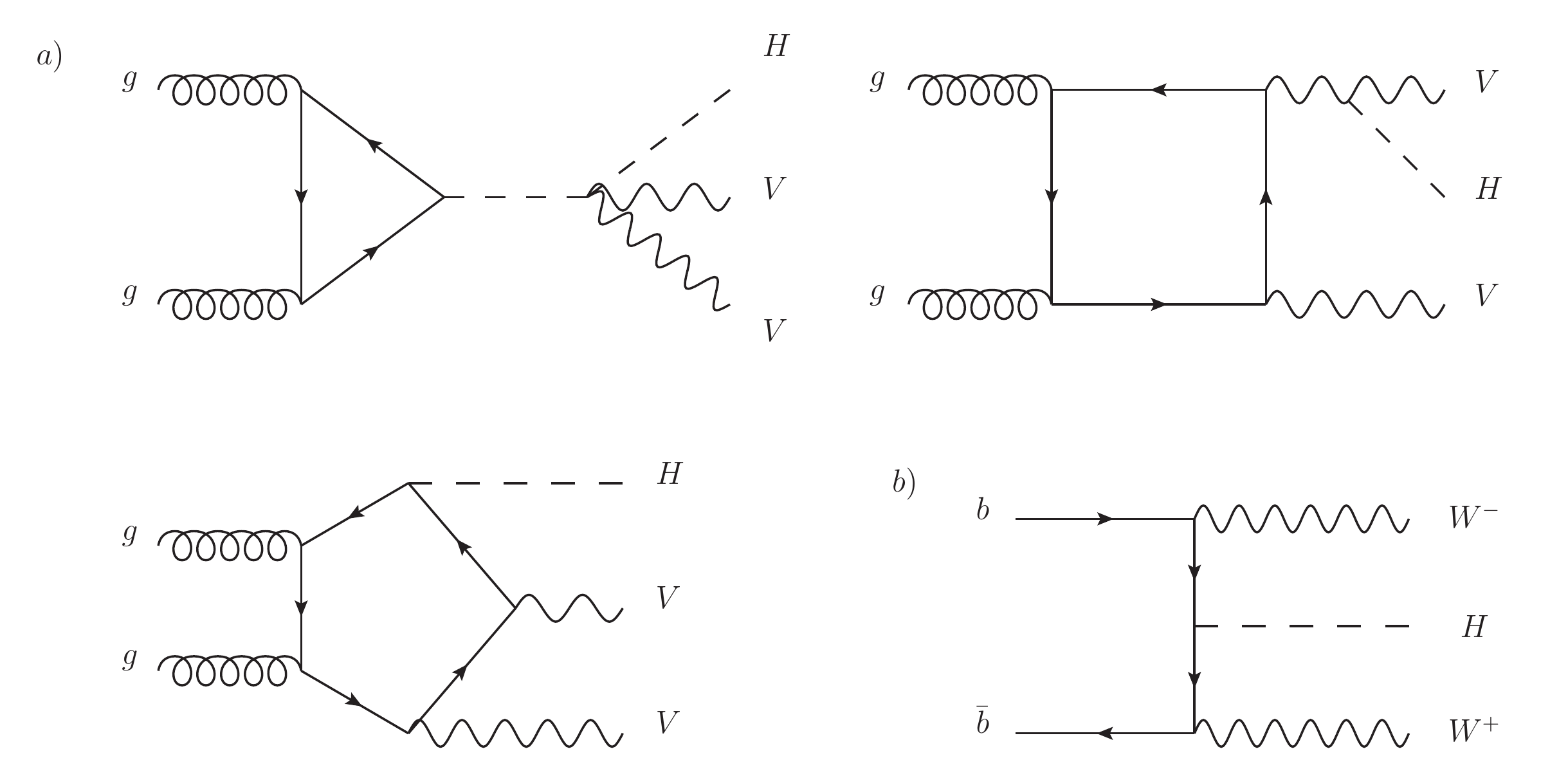}
  \caption{a) A selection of representative diagrams for $g g \to H V V$
    production processes. $V$ stands for $W$ or $Z$ bosons. b) LO
    diagram for the channel $b\bar{b}\to H W^+_{} W^-_{}$ with the top
    Yukawa coupling.
    \label{fig:diag-ggHVV}}
\end{figure*}

The leading order (LO) processes $p p \to H W^+_{} W^-_{}$ and $p p
\to H Z Z$, together with their NLO QCD corrections, have been
discussed in Ref.~\cite{Baglio:2015eon}. We only discuss in this Letter
the LO gluon fusion correction $p p\to g g\to H V V$ and the NLO
QCD-corrected bottom-quark contribution $p p\to b\bar{b}\to H V V$ at
proton-proton colliders, where $V V$ stands either for $W^+_{} W^-_{}$
or for $Z Z$. The gluon fusion channel is formally a
next-to-next-to-leading order contribution to the full hadronic cross
section $p p\to H V V$, nevertheless we will combine it with the NLO
QCD calculation of the quark channels $p p\to q \bar{q}\to H V V$
as done for the $W^+_{} W^-_{}$ and $ZZ$ production processes, see for
example Ref.~\cite{Campbell:2011bn} and references therein. The gluon
fusion correction for the $H W^+ W^-$ process was calculated in
Ref.~\cite{Mao:2009jp} and it was shown that it has an impact of
$\sim + 4\,\%$ at $M_H^{} = 120$~GeV for a central scale $\mu = (M_H^{}
+ 2 M_W^{})/2$. The $b\bar{b}$ channel has never been considered in
the literature so far.

The $gg\to H V V$ contribution is a one-loop contribution already at
the lowest order and proportional to $\alpha_s^2$. It consists of
triangle, box and pentagon loops of quarks. Our calculation is done
with five active massless flavours for the running of the strong
coupling constant $\alpha_s^{}$, and we use diagonal
Cabibbo-Kobayashi-Maskawa (CKM) matrix elements for the $H W^+_{}
W^-_{}$ process. Diagrams involving a Yukawa coupling between a light
quark and a Higgs boson are discarded, so that in the pentagon loops
only the top quark contributes. We use the 't~Hooft-Feynman gauge. We
depict in Fig.~\ref{fig:diag-ggHVV} a) some generic diagrams, in
particular the pentagon class involving the quark-quark-Higgs coupling.

The full one-loop amplitude is ultraviolet (UV) and infrared (IR)
finite and is convoluted with the gluon parton distribution functions
(PDF) to obtain the hadronic cross section as
\begin{align}
\sigma^{gg} &=& \int \dd x_1^{}\dd x_2^{}[g(x_1^{}, \mu_F^{})g(x_2^{},
\mu_F^{})\hsigma^{gg\to H VV}_{}]\, ,
\label{xsection_gg}
\end{align}
where $g(x,\mu_F^{})$ denotes the gluon PDF with momentum fraction $x$
and factorisation scale $\mu_F^{}$. The PDF evolution is taken at NLO
as well as the running of $\alpha_s^{}$. The one-loop amplitude is
generated with {\tt FeynArts-3.7}~\cite{Hahn:2000kx} and calculated
with {\tt FormCalc-8.4}~\cite{Hahn:1998yk}. The scalar integrals as
well as the reduction of tensor coefficients down to scalar integrals
are calculated using the techniques developed in
Refs.~\cite{Denner:2002ii,Denner:2005nn,Denner:2010tr} and implemented
with {\tt Collier 1.0}~\cite{Denner:2016kdg}, adapted to the {\tt
  FormCalc} framework thanks to an in-house routine. The final code is
implemented in the framework of the {\tt
  POWHEG-BOX}~\cite{Alioli:2010xd}. To improve the stability of the
calculation of the amplitudes a technical cut has been implemented,
\begin{align}
\begin{split}
  k_{i j}^{}\geq k_{\rm cut}^{},\ \ \  k_{i j} = {\rm min}(\tilde{k}_{i j}^{},
  p_{T,i}^{}, p_{T,j}^{}),\\
  \tilde{k}_{i j}^{} = \frac35 {\rm min}(p_{T,i}^{}, p_{T,j}^{})
  \sqrt{\Delta y_{i j}^2 + \Delta \phi_{i j}^2},
\end{split}
\end{align}
where $(ij)$ runs on the pairs of final-state particles ($i\neq j$),
$\Delta y_{i j}$ and $\Delta \phi_{i j}$ are the rapidity and angular
separations between the particle $i$ and $j$, $p_{T,i}$ is the
transverse momentum of particle $i$, and $k_{\rm cut}^{} =
10^{-2}_{}$. This technical cut helps to get rid of regions where the
Gram determinant of the tensor integrals is close to zero, acting in
much the same way as a jet veto. It has been checked that the result
does not depend on the value of $k_{\rm cut}^{}$ (as long as $k_{\rm
  cut}^{}$ is small enough). Note that the calculation could be done
without this cut, but at the cost of increasing the number of points
in the integration routine, thus slowing down the whole calculation.

\begin{figure*}[t]
   \centering
   \includegraphics[scale=0.72]{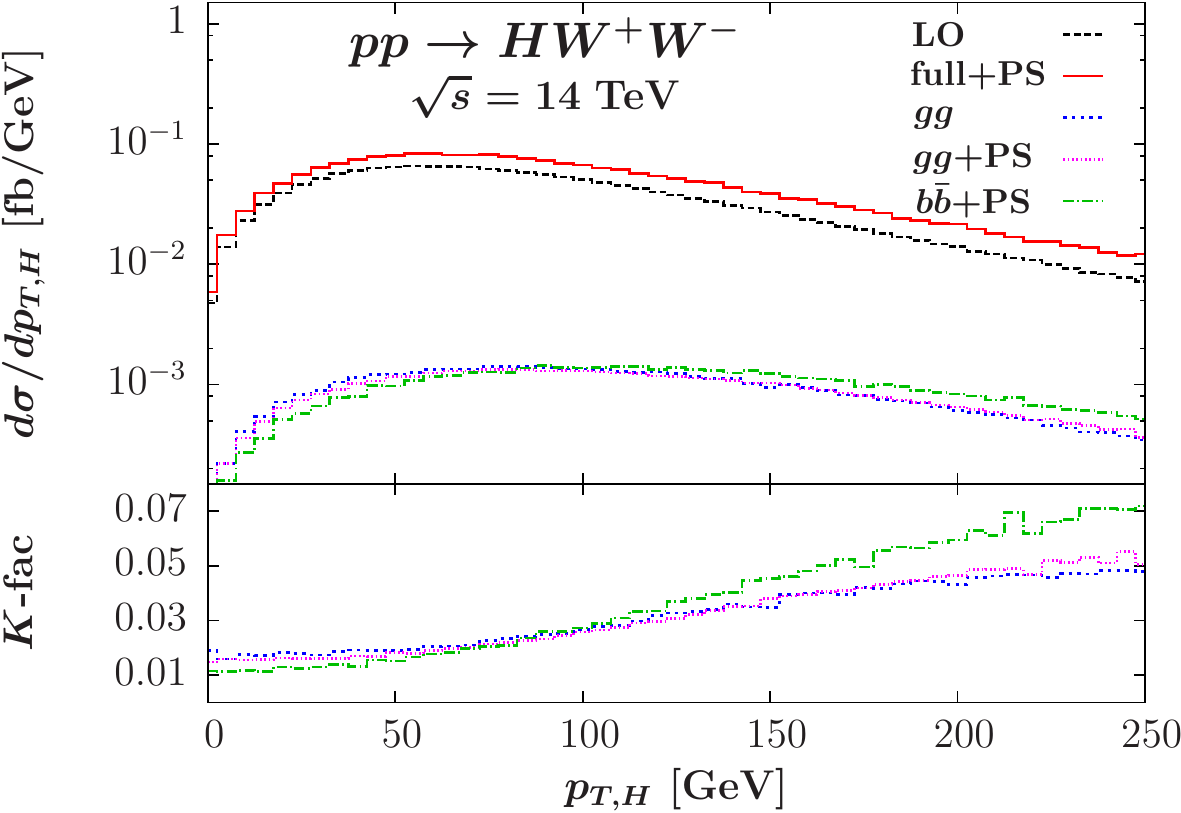}
   \hspace{6mm}
   \includegraphics[scale=0.72]{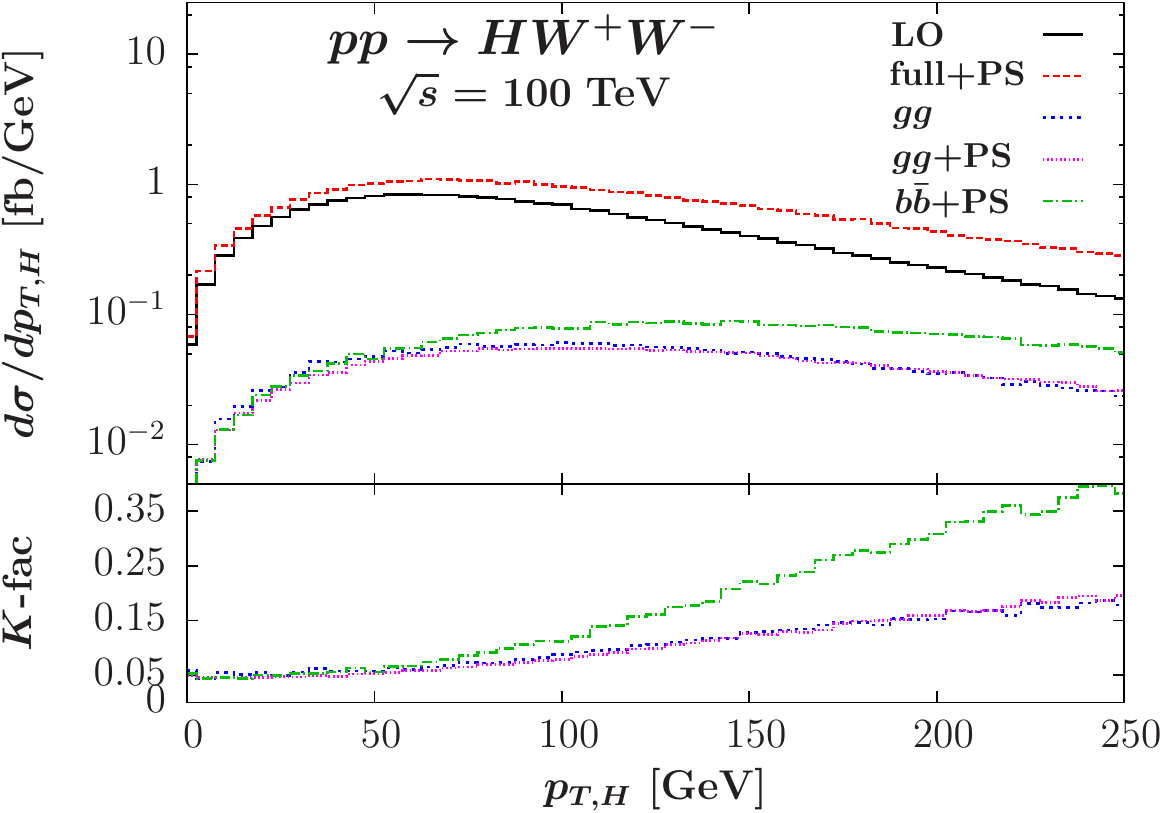}
   \caption{In the main frame: Higgs transverse momentum $p_{T,H}^{}$
     (in GeV) distribution of the $p p \to H W^+_{} W^-_{}$ cross
     section (in fb/GeV) at the 14 TeV LHC (left) and at the 100 TeV
     FCC-hh (right) calculated with the {\tt PDF4LHC15\_nlo} PDF set
     and with the input parameters given in
     Eq.~(\ref{param-setup}). In black (dashed): the LO QCD
     distribution; in red (solid): the full distribution including NLO
     QCD corrections as well as the $g g$ and $b\bar{b}$
     contributions, corrected with PS effects; in blue (dotted): the
     $g g$ contribution; in pink (thin dotted): the $g g$ contribution
     including PS effects; in green (dash-dotted): the $b\bar{b}$
     contribution including PS effects. In the insert are displayed
     the $g g$, the $g g$+PS and the $b\bar{b}$+PS $K$--factors
     relative to the LO prediction. The $b\bar{b}$ contribution
     without PS effects would be nearly the same as the curve
     including them.
     \label{fig:HWW-pTH-dist}}
 \end{figure*}

The bottom-quark contribution in the case of the $H Z Z$ channel is
calculated in much the same way as in Ref.~\cite{Baglio:2015eon} for
the light quark contributions, as we take the bottom quark massless
and use the bottom-quark PDF in a five-flavour-scheme. The Feynman
diagrams are similar and the renormalisation is the same. The
calculation of the channel $p p\to b\bar{b} \to H W^+_{} W^-_{}$
includes one new diagram at LO, depicted in Fig.~\ref{fig:diag-ggHVV}
b), involving the top-quark Yukawa coupling.

The real corrections at NLO in the $H W^+_{} W^-_{}$ channel
involve resonant diagrams with on-shell top-quark which lead to a 
double-counting with the production process $p p\to H W^-_{} t$
followed by the on-shell decay $t\to W^+_{} b$. This issue is well
known in $W$-pair production (see for example
Refs.~\cite{Bierweiler:2012kw,Baglio:2013toa}) and one way to solve
it is to introduce a $b$-jet veto with a $100\,\%$ efficiency as
done in Ref.~\cite{Bierweiler:2012kw} for $W$-pair production. If a
more realistic jet veto is wanted, a more complicated diagram
subtraction method has to be used~\cite{Gehrmann:2014fva}. We will
not enter in such details which go beyond the scope of this paper as
it would not change significantly the amount of QCD corrections to
the bottom-quark fusion process, which itself is a sub-leading
channel of the full process $p p\to H W^+_{} W^-_{}$. To implement
this $b$-jet veto we simply omit the contributions with a gluon in
the initial state at the level of the PDFs.

We use the same phase-space parametrisation that was used in our
previous work~\cite{Baglio:2015eon} for both gluon fusion and
bottom-quark fusion contributions. It has been checked explicitly that
the amplitudes are UV and IR finites. In addition, adopting the same
framework as in Ref.~\cite{Mao:2009jp} for the $H W^+_{} W^-_{}$
process, a good agreement has been found between their results and our
calculation.

\section{Numerical results at the LHC and at the FCC}
\label{sec:numres}

Following strictly the framework of our previous\linebreak
work~\cite{Baglio:2015eon}, we use the following set of input
parameters,
\begin{align}
\begin{split}
G_F^{} & = 1.16637\times 10^{-5}_{} \text{ GeV}^{-2}_{}, \,\, 
M_W^{}=80.385 \text{ GeV},\\
 M_Z^{} & = 91.1876 \text{ GeV},\,\, M_t^{} = 172.5 \text{ GeV},\\
 M_H^{} & =125 \text{ GeV}, \,\, \alpha_s^{\rm NLO}(M_Z^2)= 0.118,
\end{split}
\label{param-setup}
\end{align}
where all but $M_H^{}$ is taken from Ref.~\cite{Agashe:2014kda}. The
CKM matrix is assumed to be diagonal and the masses of all the quarks
but the top quark are approximated as zero. Following the latest
PDF4LHC Recommendation~\cite{Butterworth:2015oua} we use in the {\tt
  LHAPDF6} framework~\cite{Buckley:2014ana} the NLO PDF set family
{\tt PDF4LHC15\_nlo} which combines the three global sets {\tt
  CT14}~\cite{Dulat:2015mca}, {\tt MMHT14}~\cite{Harland-Lang:2014zoa}
and {\tt NNPDF3.0}~\cite{Ball:2014uwa} using the combination method
developed in~\cite{Gao:2013bia,Carrazza:2015aoa}. To define the jets
we use {\tt FastJet}~\cite{Cacciari:2005hq,Cacciari:2011ma} and the
parton shower is done with {\tt Pythia
  6.4}~\cite{Sjostrand:2006za}. The central scale is defined as the $H
V V$ invariant mass, $\mu_R^{} = \mu_F^{} = \mu_0^{}$ with
$\mu_0^{HWW} = M_{HW^+W^-}^{}$, $\mu_0^{HZZ} = M_{HZZ}^{}$. The
running of $\alpha_s^{}$ is done at NLO throughout the whole Letter.

Using this set of parameters we obtain at the LHC at 14~TeV a $\sim
+3\,\%$ correction to the LO $p p\to H W^+_{} W^-_{}$ cross section
and a $\sim +5\,\%$ correction to the LO $p p\to H Z Z$ cross section
coming from the $gg$ contributions. At the FCC-hh at 100~TeV we obtain
$\sim +10\,\%$ and $\sim +18\,\%$ corrections respectively. The
correction increases, in both channels, with increasing centre-of-mass
(c.m.) energies. The $b\bar{b}$ contributions are less important for
the $HZZ$ channel, leading to a $\sim +2\,\%$ increase at the LHC and
a $\sim +7\,\%$ increase at the FCC-hh. The NLO QCD corrections to the
$b\bar{b}$ channel itself are very small, at most $\sim +3\,\%$ at the
FCC-hh. In contrast, the impact of the $b\bar{b}$ channel is
significant in the $H W^+_{} W^-_{}$ channel already at LO. The new
$t$--channel diagram depicted in Fig.~\ref{fig:diag-ggHVV}, together
with the other diagrams belonging to the same gauge-invariant class,
induces this significant contribution, in particular at the
FCC-hh. The NLO QCD corrections to $b\bar{b}\to H W^+_{} W^-_{}$ are
small, $\sim +4\,\% (+0.3\,\%)$ at the LHC (FCC-hh). The correction
from this channel to the LO $p p\to q\bar{q}\to H W^+_{} W^-_{}$
process is $\sim +3\,\%$ at the LHC and $\sim +18\,\%$ at the
FCC-hh. The total QCD corrections, including the NLO contributions
calculated in Ref.~\cite{Baglio:2015eon}, eventually amount to $\sim
+33\,\%$ ($\sim +30\,\%$) for the $H W^+_{} W^-_{}$ ($H Z Z$) channel
at the LHC and to $\sim +57\,\%$ ($\sim +42\,\%$) for the $H W^+_{}
W^-_{}$ ($H Z Z$) channel at the FCC-hh.

\begin{figure*}[t]
   \centering
   \includegraphics[scale=0.72]{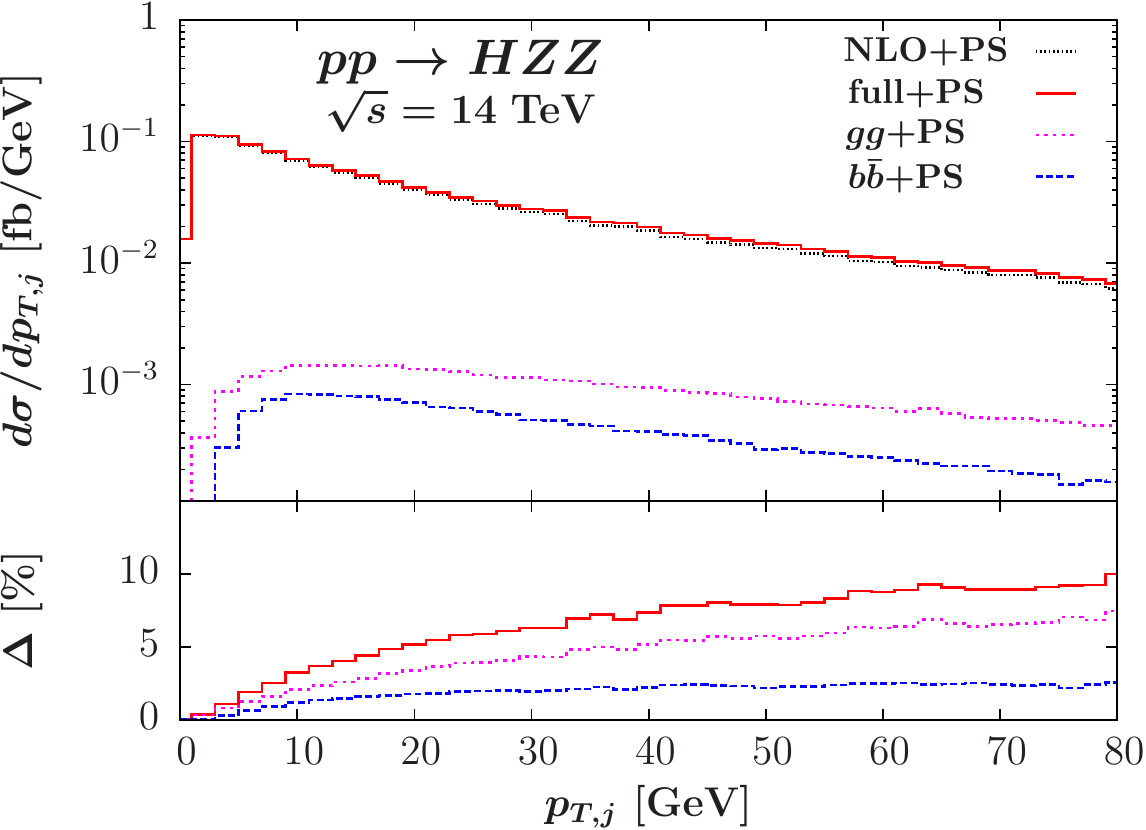}
   \hspace{6mm}
   \includegraphics[scale=0.72]{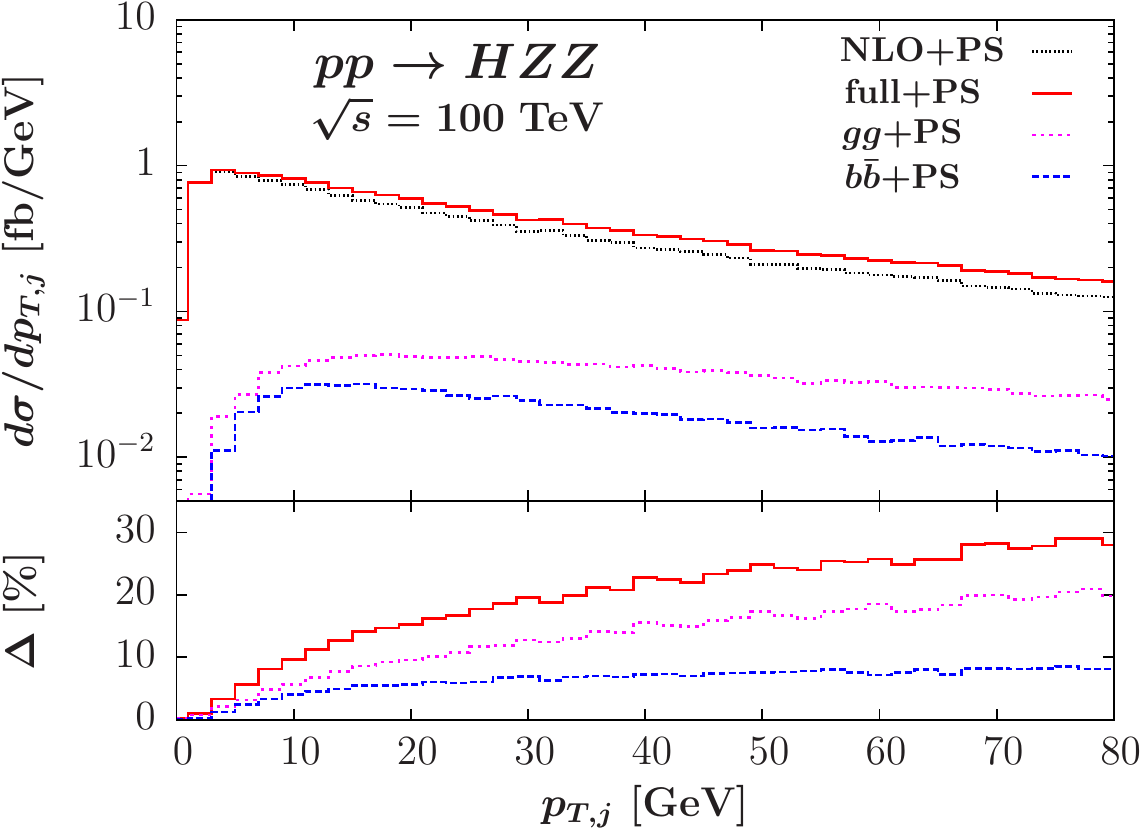}
   \caption{In the main frame: Jet transverse momentum $p_{T,j}^{}$
     (in GeV) distribution of the $p p \to H Z Z$ cross section (in
     fb/GeV) at the 14 TeV LHC (left) and at the 100 TeV FCC-hh
     (right) calculated with the {\tt PDF4LHC15\_nlo} PDF set and with
     the  input parameters given in Eq.~(\ref{param-setup}). In black
     (thin dotted): the NLO distribution including PS effects; in red
     (solid): the full distribution including NLO QCD corrections as
     well the $g g$ and $b\bar{b}$ contributions, corrected  with PS
     effects; in pink (dotted): the $g g$ contribution including PS
     effects; in blue (dashed): the $b\bar{b}$ contribution including
     PS effects. In the insert are displayed the corrections $\Delta$
     (in \%) due to the gluon fusion and $b\bar{b}$ contributions (in
     pink/dotted and blue/dashed respectively) as well as their sum
     (in red/solid) with respect to the NLO+PS calculation.
     \label{fig:HZZ-pTj-dist}}
 \end{figure*}

\subsection{Differential distributions}
\label{sec:numres:diff}

As an example of the impact of the gluon fusion and $b\bar{b}$
corrections on the differential distributions we present the Higgs
transverse momentum $p_{T,H}$ for the $H W^+_{} W^-_{}$ channel and
the jet transverse momentum $p_{T,j}$ distributions for the $H Z Z$
channel, in both cases at the LHC and at the FCC-hh. The other
kinematic distributions can be easily obtained in the {\tt POWHEG-BOX}
framework. Note that the distributions are quite similar between both
channels, so that for example the conclusions drawn in the $H W^+_{}
W^-_{}$ for the $p_{T,H}$ distribution apply also in the $H Z Z$
channel, but as expected from the corrections on the total cross
sections the impact of gluon fusion contributions is more visible in
the $H Z Z$ channel than in the $H W^+_{} W^-_{}$ channel and
vice-versa for the impact of $b\bar{b}$ contributions.

In Figure~\ref{fig:HWW-pTH-dist} we display the $p_{T,H}$
distributions at the LHC (left) and the FCC-hh (right), including the 
LO prediction (in black/dashed line) and the full prediction including
both the NLO QCD corrections and the gluon fusion and $b\bar{b}$
contributions (in red/solid line), as well as the gluon fusion
contribution alone (in blue/dotted line) and with parton shower (PS)
effects simulated with {\tt Pythia} (in pink/thin dotted line), and
finally the $b\bar{b}$ contribution including PS effects (in
green/dash-dotted line). The contribution from $b\bar{b}$ subprocess
alone, without PS effects, would be nearly identical to the curve
including PS effects. The inserts display the $K$-factor of the gluon
fusion and $b\bar{b}$ contributions, defined as $K=\sigma^{g
  g/b\bar{b}}_{}/\sigma^{\rm LO}_{}$. The $g g$ corrections are quite
small at the LHC, of the order of $+2\,\%$ to $+5\,\%$, linearly
increasing from low to high $p_{T,H}^{}$, but sizeable at the FCC-hh,
from $+5\,\%$ to $\sim +18\,\%$. The PS effects are quite small on the
$g g$ contribution but noticeable as a change in shape, leading to
slightly smaller corrections at low $p_{T,H}$ and slightly larger
corrections at high $p_{T,H}$. The $b\bar{b}$ effects follow the same
pattern, except that the PS effects are not significant and that the
shape of the $b\bar{b}$ corrections is more quadratic. At the LHC they
are of the order of $+1\,\%$ to $+7\,\%$, smaller than $gg$
corrections at low $p_{T,H}^{}$ and bigger at high transverse
momenta. The corrections at the FCC-hh are important, reaching $\sim
+40\,\%$ at $p_{T,H}^{} = 250$~GeV.

In Figure~\ref{fig:HZZ-pTj-dist} we display the $p_{T,j}$
distributions at the LHC (left) and the FCC-hh (right), including the
PS effects everywhere. The NLO distribution is displayed in black (thin
dotted line), the full distribution including the $g g$ and $b\bar{b}$
contributions is displayed in red (solid line), and the $g g$
contribution as well as the $b\bar{b}$ contribution, alone, are
displayed in pink (dotted line) and blue (dashed line)
respectively. The insert displays the percent correction due to the $g
g$ contribution, the $b\bar{b}$ contribution as well as the sum of
them, with respect to the NLO prediction. The $g g$ contribution
displays a logarithmic increase with increasing $p_{T,H}$, and is
modest at the LHC but sizeable at the FCC-hh. The $b\bar{b}$
contributions are negligible for low $p_{T,j}^{}$ and then are flat,
of the order of $\sim + 2.5\,\%$ at the LHC and $\sim + 9\,\%$ at the
FCC-hh. The sum of the two contributions reach, at high $p_{T,j}^{}$,
$+10\,\%$ at the LHC and $+30\,\%$ at the FCC-hh.

\subsection{Total cross sections including theoretical uncertainties}
\label{sec:numres:total}

As already demonstrated in Ref.~\cite{Baglio:2015eon} the total rates
$p p\to H W^+_{} W^-_{}/H Z Z$ are affected by theoretical
uncertainties: 1) the scale uncertainty reflecting the confidence
given to the calculation at a given perturbative order, calculated by
varying the renormalisation scale $\mu_R$ and the factorisation scale
$\mu_F$ in the range $\displaystyle \frac12 \mu_0^{} \leq
\mu_R^{},\mu_F^{}\leq 2\mu_0^{}$; 2) the PDF+$\alpha_s^{}$ uncertainty
reflecting the impact of the experimental uncertainties on the fit
leading to the determination of the PDFs. We calculate the theoretical
uncertainties on the gluon fusion and the $b\bar{b}$ contribution as
well as on their combination with the NLO QCD $q\bar{q}$ contribution
to the whole hadronic cross section, following
Ref.~\cite{Baglio:2015eon}.

\begin{figure}
   \centering
   \includegraphics[scale=0.8]{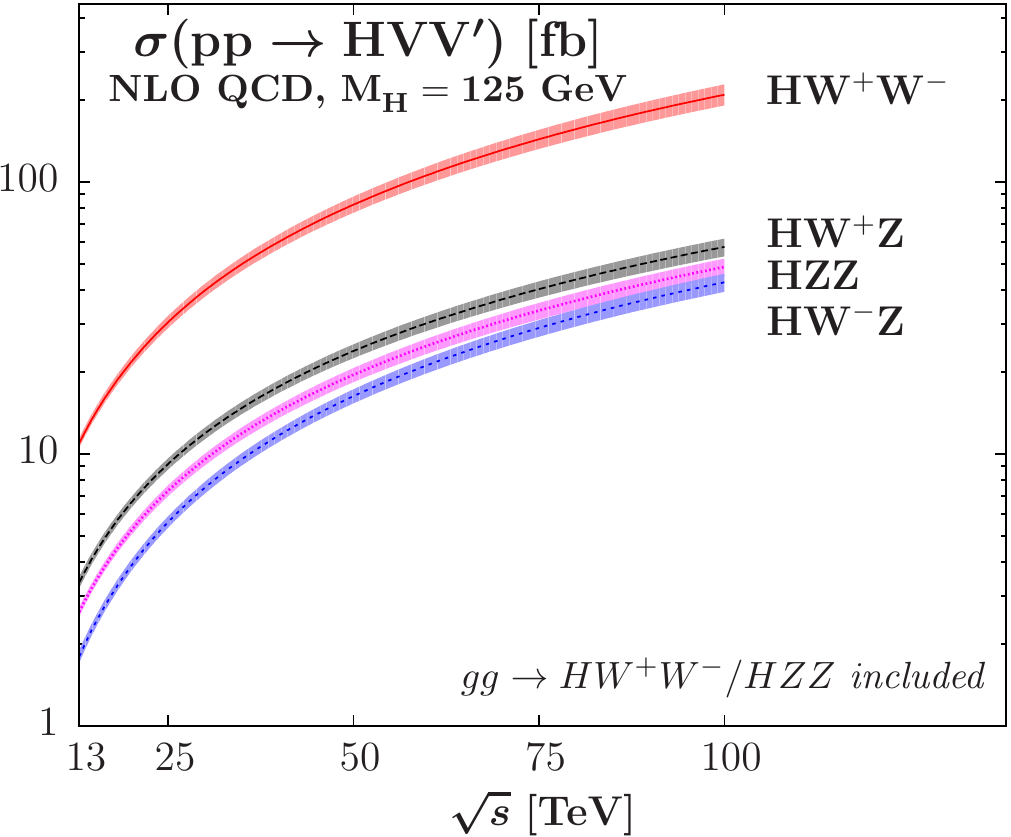}
   \caption{The total cross sections (in fb) for SM Higgs production
     in association with a pair of weak bosons at NLO QCD as a
     function of the c.m. energy (in TeV) with $M_H^{}=125$ GeV:
     $HW^+_{} W^-_{}$ (red/full), $HW^+_{} Z$ (grey/dashed), $HW^-_{}
     Z$ (pink/dotted) and $HZZ$ (blue/dashed with small dashes). The
     gluon fusion and $b\bar{b}$ contributions are included in the
     $HW^+_{} W^-_{}$ and $H Z Z$ production channels. The {\tt
       PDF4LHC2015\_30} PDF set has been used and the total
     theoretical uncertainties are included as corresponding bands
     around the central values. The data for the $W Z$ channels comes
     from Ref.~\cite{Baglio:2015eon}.
     \label{fig:totalxsfinal}}
 \end{figure}

The results are presented in Tables~\ref{tab:HWWtotal} and
~\ref{tab:HZZtotal} as well as displayed in
Figure~\ref{fig:totalxsfinal}. The scale uncertainty of the gluon
fusion cross sections is quite large, of the order of $15\,\%$ to
$30\,\%$. This was expected as this is a LO process. The scale
uncertainties of the subprocess $b\bar{b}\to H W^+_{} W^-_{}$ are also
quite large, due to the $b$--jet veto that we use and is mainly driven
by the variation of the factorisation scale. We have checked that if a
similar jet veto were used for the light-quark contributions the scale
uncertainty would also rise and reach for example $\sim \pm 10\,\%$ at
the FCC-hh instead of the $\sim \pm 4\,\%$ uncertainty quoted in
Ref.~\cite{Baglio:2015eon}. In the case of the $b\bar{b}\to H Z Z$
subprocess the scale uncertainty is slightly larger that in
Ref.~\cite{Baglio:2015eon} for the light quark contributions.

The combination of the gluon fusion and $b\bar{b}$ channels
with the NLO $q\bar{q}$ cross section, however, displays limited
uncertainties albeit larger than for the $q\bar{q}$ channel alone,
because of the sizeable impact of the $g g$ correction in the $H Z Z$
channel and the $b\bar{b}$ correction in the $H W^+_{} W^-_{}$
channel, in particular at the FCC-hh. The PDF+$\alpha_s$ uncertainty
is nearly the same in the $q\bar{q}$ contributions and in the full
cross sections. We obtain a total theoretical uncertainty of $\sim \pm
4\,\% (\sim \pm 9\,\%)$ at the LHC (FCC-hh) for the $H W^+_{} W^-_{}$
channel, compared to $\sim \pm 4\,\% (\sim +6\,\% / -7\,\%)$ for the
NLO QCD $q\bar{q}$ contributions only~\cite{Baglio:2015eon}.  In the
case of the $H Z Z$ channel we obtain a total theoretical uncertainty
of $\sim \pm 4\,\% (+7\,\% / -8\,\%)$ at the LHC (FCC-hh), compared to
$\sim \pm 4\,\% (+5\,\% / -7\,\%)$ for the NLO QCD $q\bar{q}$
contributions only~\cite{Baglio:2015eon}. The impact of the gluon
fusion and $b\bar{b}$ contributions uncertainties are then negligible
at LHC energies but noticeable at the FCC-hh. Note that in comparison
with Ref.~\cite{Baglio:2015eon} the $H Z Z$ channel dominates over the
$H W^-_{} Z$ channel not only at lower c.m. energies but also at
higher c.m. energies, when the gluon fusion and the $b\bar{b}$
corrections are included.

\begin{table}
  \renewcommand{\arraystretch}{1.5}
  \centering
  \caption{The total $HW^+_{} W^-_{}$ production cross
    section at the LHC and at the FCC-hh (in fb) for given c.m. energies
    (in TeV) at the central scale $\mu_F^{} = \mu_R^{} =
    M_{HWW}^{}$. The first group of lines displays the $g g\to H
    W^+_{} W^-_{}$ contribution, the second group displays the
    $b\bar{b}\to H W^+_{} W^-_{}$ contribution and the third displays the
    combination of these contributions with the NLO QCD $q\bar{q}$
    contribution taken from Ref.~\cite{Baglio:2015eon}. The
    corresponding shifts due to the theoretical uncertainties coming
    from scale variation, PDF+$\alpha_s^{}$ errors as well as the
    total uncertainty, when all errors are added linearly, are also
    shown (in \%).\smallskip}
  \begin{tabular}{c|cccc}
    \hline
    $\sqrt{s}$ [TeV]
    & $\sigma^{g g}_{HWW}$ [fb]
    & Scale & PDF$+\alpha_s^{}$ & Total\\
    \hline
    13 & 0.217 & $^{+27\,\%}_{-21\,\%}$ & $^{+4.2\,\%}_{-4.2\,\%}$ & $^{+31\,\%}_{-25\,\%}$\\
    14 & 0.262 & $^{+26\,\%}_{-20\,\%}$ & $^{+3.5\,\%}_{-3.5\,\%}$ & $^{+30\,\%}_{-24\,\%}$\\
    33 & 1.81 & $^{+21\,\%}_{-16\,\%}$ & $^{+2.4\,\%}_{-2.4\,\%}$ & $^{+23\,\%}_{-18\,\%}$\\
    100 & 13.8 & $^{+22\,\%}_{-18\,\%}$ & $^{+2.2\,\%}_{-2.2\,\%}$ & $^{+24\,\%}_{-20\,\%}$\\
    \hline
    $\sqrt{s}$ [TeV]
    & $\sigma^{b\bar{b}}_{HWW}$ [fb]
    & Scale & PDF$+\alpha_s^{}$ & Total\\
    \hline
    13 & 0.236 & $^{+22\,\%}_{-18\,\%}$ & $^{+7.8\,\%}_{-7.8\,\%}$ & $^{+30\,\%}_{-26\,\%}$\\
    14 & 0.293 & $^{+22\,\%}_{-18\,\%}$ & $^{+7.6\,\%}_{-7.6\,\%}$ & $^{+30\,\%}_{-25\,\%}$\\
    33 & 2.63 & $^{+24\,\%}_{-19\,\%}$ & $^{+5.7\,\%}_{-5.7\,\%}$ & $^{+29\,\%}_{-24\,\%}$\\
    100 & 24.5 & $^{+27\,\%}_{-20\,\%}$ & $^{+4.1\,\%}_{-4.1\,\%}$ & $^{+31\,\%}_{-24\,\%}$\\
    \hline
    $\sqrt{s}$ [TeV]
    & $\sigma^{\rm full}_{HWW}$ [fb]
    & Scale & PDF$+\alpha_s^{}$ & Total\\
    \hline
    13 & 11.0 & $^{+2.7\,\%}_{-1.9\,\%}$ & $^{+1.7\%}_{-1.7\%}$ & $^{+4.4\,\%}_{-3.7\,\%}$\\
    14 & 12.4 & $^{+2.8\,\%}_{-2.1\,\%}$ & $^{+1.7\%}_{-1.7\%}$ & $^{+4.5\,\%}_{-3.8\,\%}$\\
    33 & 45.9 & $^{+4.5\,\%}_{-4.0\,\%}$ & $^{+1.5\%}_{-1.5\%}$ & $^{+6.0\,\%}_{-5.5\,\%}$\\
    100 & 209 & $^{+7.2\,\%}_{-6.9\,\%}$ & $^{+1.9\%}_{-1.9\%}$ & $^{+9.1\,\%}_{-8.7\,\%}$\\
    \hline
  \end{tabular}
   \label{tab:HWWtotal}
 \end{table}

\begin{table}
 \renewcommand{\arraystretch}{1.5}
  \centering
   \caption{Same as Table~\ref{tab:HWWtotal} but for the $HZZ$ channel
     at the central scale $\mu_F^{} = \mu_R^{} =
     M_{HZZ}^{}$.\smallskip}
  \begin{tabular}{c|cccc}
    \hline
    $\sqrt{s}$ [TeV]
    & $\sigma^{g g}_{HZZ}$ [fb]
    & Scale & PDF$+\alpha_s^{}$ & Total\\
    \hline
    13 & 0.093 & $^{+27\,\%}_{-20\,\%}$ & $^{+3.2\,\%}_{-3.2\,\%}$ & $^{+31\,\%}_{-23\,\%}$\\
    14 & 0.113 & $^{+27\,\%}_{-20\,\%}$ & $^{+3.2\,\%}_{-3.2\,\%}$ & $^{+30\,\%}_{-23\,\%}$\\
    33 & 0.783 & $^{+20\,\%}_{-16\,\%}$ & $^{+2.2\,\%}_{-2.2\,\%}$ & $^{+22\,\%}_{-18\,\%}$\\
    100 & 6.02 & $^{+22\,\%}_{-18\,\%}$ & $^{+2.1\,\%}_{-2.1\,\%}$ & $^{+24\,\%}_{-20\,\%}$\\
    \hline
    $\sqrt{s}$ [TeV]
    & $\sigma^{b\bar{b}}_{HZZ}$ [fb]
    & Scale & PDF$+\alpha_s^{}$ & Total\\
    \hline
    13 & 0.036 & $^{+4.5\,\%}_{-4.7\,\%}$ & $^{+8.0\,\%}_{-8.0\,\%}$ & $^{+13\,\%}_{-13\,\%}$\\
    14 & 0.044 & $^{+4.4\,\%}_{-4.8\,\%}$ & $^{+7.8\,\%}_{-7.8\,\%}$ & $^{+12\,\%}_{-13\,\%}$\\
    33 & 0.323 & $^{+4.2\,\%}_{-5.2\,\%}$ & $^{+5.9\,\%}_{-5.9\,\%}$ & $^{+10\,\%}_{-11\,\%}$\\
    100 & 2.49 & $^{+3.7\,\%}_{-6.5\,\%}$ & $^{+4.8\,\%}_{-4.8\,\%}$ & $^{+8.5\,\%}_{-11\,\%}$\\
    \hline
    $\sqrt{s}$ [TeV]
    & $\sigma^{\rm full}_{HZZ}$ [fb]
    & Scale & PDF$+\alpha_s^{}$ & Total\\
    \hline
    13 & 2.63 & $^{+2.4\,\%}_{-1.8\,\%}$ & $^{+1.8\,\%}_{-1.8\,\%}$ & $^{+4.2\,\%}_{-3.7\,\%}$\\
    14 & 2.98 & $^{+2.4\,\%}_{-2.0\,\%}$ & $^{+1.7\,\%}_{-1.7\,\%}$ & $^{+4.2\,\%}_{-3.7\,\%}$\\
    33 & 11.0 & $^{+3.4\,\%}_{-3.5\,\%}$ & $^{+1.6\,\%}_{-1.6\,\%}$ & $^{+5.1\,\%}_{-5.1\,\%}$\\
    100 & 48.6 & $^{+5.1\,\%}_{-5.9\,\%}$ & $^{+2.1\,\%}_{-2.1\,\%}$ & $^{+7.2\,\%}_{-8.0\,\%}$\\
    \hline
  \end{tabular}
   \label{tab:HZZtotal}
 \end{table}


\section{Conclusion}
\label{sec:conclusions}

We have completed in this Letter the current picture of the QCD
corrections to $H W^+_{} W^-_{}$ and $H Z Z$ productions at hadron
colliders in the {\tt POWHEG-BOX} framework, including the matching
with parton shower, by calculating the gluon fusion corrections $g
g\to H W^+_{}W^-_{} / H Z Z$ and the bottom quark fusion corrections
$b\bar{b}\to H W^+_{} W^-_{}$. The latter are studied in this Letter
for the first time and their NLO QCD corrections are also
included. The gluon fusion contributions are sizeable, from $\sim
+5\,\%$ at LHC energies up to $\sim +18\,\%$ at 100 TeV. The
$b\bar{b}$ contributions are particularly important at the FCC-hh in
the $H W^+_{} W^-_{}$ channel, where they reach $\sim
+18\,\%$. Combining the NLO corrections to the $p p\to q\bar{q}\to H
W^+_{} W^-_{} / H Z Z$ cross sections already calculated in
Ref.~\cite{Baglio:2015eon} with the two new contributions studied in
this Letter, the QCD corrections amounts to $+33\,\%$ $(+57\,\%)$ and
$+30\,\%$ $(+42\,\%)$ for the total cross sections $p p\to H W^+_{}
W^-_{}$ and $p p\to H Z Z$ at the LHC (FCC-hh), respectively. The
total theoretical uncertainty is nearly unmodified at LHC energies
while the change is noticeable at the FCC-hh, from $\sim +6\,\%
/ - 7\,\%$ to $\sim \pm 9\,\%$ for the $H W^+_{} W^-_{}$ channel and
from $\sim +5\,\% / - 7\,\%$ to $\sim +7\,\% / -8\,\%$ for the $H Z Z$
channel. The impact of the gluon fusion and $b\bar{b}$ channels on the
differential distributions matched to parton shower is found small at
the LHC and sizeable at the FCC-hh, following the pattern of the
corrections on the total cross sections. A public release of the code
in the {\tt POWHEG-BOX} is expected in the near future.\newline

\noindent{\bf Acknowledgments}\newline

The author thanks Juraj Streicher and Nad\`ege Bernard for their
reading of the manuscript as well as Valentin Hirschi and Barbara
J\"ager for discussions. The referee is also aknowledged for his
suggestion to study the $b\bar{b}$ contributions. This work was
performed thanks to the support of the state of Baden-W\"{u}rttemberg
through bwHPC and the German Research Foundation (DFG) through grant
no INST 39/963-1 FUGG. The author was supported in part by the
Institutional Strategy of the University of T\"{u}bingen (DFG, ZUK 63)
and the DFG Grant JA 1954/1.\newline\newline

\bibliography{ggHVV_paper_litt}

\end{document}